\documentstyle[twocolumn,amssymb,aps,epsfig]{revtex}
\begin{document}

\title{\bf Pure Spin Currents and associated electrical voltage}
\author{T.~P.~Pareek}
\address{
Max-Planck-Institute f\"ur Mikrostrukturphysik, Weinberg 2,
D-06120 Halle, Germany}
\address{~
\parbox{14cm}{\rm
\medskip
We present a generalize Landauer-B\"uttiker transport theory for
multi-terminal spin transport in presence of spin-orbit interaction. 
It is pointed out that the presence of
spin-orbit interaction results in {\it equilibrium spin currents},
since in presence of spin-orbit interaction spin is not a
conserved quantitative.
Further  we illustrate the theory by applying it to a  three terminal
Y-shaped conductor. It is shown that when one of the terminal is
a potential probe, there exist {\it nonequilibrium pure spin currents} without an
accompanying charge current. It is shown that this 
pure spin currents causes a voltage drop which can be measured if
the
potential probe is  ferromagnetic.
}}
\maketitle
\narrowtext

Producing and measuring {\it spin  currents} is  a major goal of
spintronics. The standard way is to inject spin currents from a
Ferromagnet into a semiconductor in a two terminal
geometry \cite{wolf}. 
However this has a drawback, due to conductivity mismatch, the
polarization of injected current is rather small and it always has
an accompanying charge current \cite{molenkamp}. Also for any spintronics operation
spin orbit interaction plays an important role, for e.g., in Datta-Das
spin-transistor \cite{datta}. 

In light of these development
it would be interesting and highly desirable if one can produce spin
currents intrinsically. One such possibility is provided by
intrinsics spin-orbit interaction. Presence of impurity atom or
defects
gives rise to spin-orbit interaction of the form \cite{hirsch}
\cite{bergman},\cite{tribhu},
\begin{equation}
H_{so}=\lambda (\bbox{\nabla} U(r) \times \bf{k}) \cdot 
\bbox{\sigma} 
\label{eq_so}
\end{equation}
\noindent where $\bbox{\sigma}$ is a vector of Pauli matrices ,${\bf
  U}(r)$
is potential due to defects or impurity atoms and ${\bf k}$ is the
momentum wave vector of electrons and $\lambda$ is spin-orbit
interaction strength. 
 For strictly two dimensional case for which the
potential
${\bf U}(r)$ depends on {\it x} and {y} coordinates only
the Hamiltonian commutes with $\sigma_{z}$ , hence {\it z} component
of the spin is good quantum number.
As is well know that this kind of
spin-orbit interaction has a polarizing effect on particle scattering
\cite{Landau}, i.e, when  an unpolarized beam is scattered it gets
polarized perpendicular to the plane of scattering. Further
scattering of this polarized beam causes asymmetry in scattering
processes, {\it i.e.}
electrons with one particular spin direction, e.g., spin-up electrons
have a larger probability to be scattered to the right compared to
spin-down electrons \cite{tribhu},\cite{Landau}. This property of spin-orbit
scattering gives rise to novel effects like spin hall effect 
\cite{hirsch}.

Here in this article we show that the the above discussed property of
spin-orbit scattering can be used to generate and measure {\it
  spin-currents} \cite{bergman}. Consider a
three terminal, two dimensional Y shaped conductor shown in Fig.1.
The plane of conductor is {\it xy}. Since the conductor is two
dimensional which fixes the scattering plane, the scattered electrons will be polarized along
{\it z} axis (perpendicular to the scattering plane). 
However the polarization for the two branches of
Y junction will be opposite \cite{hirsch}. Hence a three terminal 
structure would create spin currents from an unpolarized current in
presence
of spin-orbit interaction \cite{tribhu}, \cite{kiseleve}.
Moreover a three terminal device provides an important
possibility of generating {\it nonequilibrium pure spin currents} without an accompanying
charge current.
This is the case  when one of the terminal acts as a voltage probe.
For e.g., say the terminal 3 is a voltage probe as shown in Fig.1,
{\it i.e.}, the voltage $V_3$ at third probe is adjusted such that
the total charge current flowing in terminal 3 is zero, {\it i.e.}
$I_{3}^q$=0 \cite{but1} ,\cite{tribhu3}. 
Physically it implies that the charge current flowing in
(which is polarized as argued above) is same as charge current flowing
out. {\it However the polarization of charge current flowing out need not to 
be same
as the polarization of charge current flowing in}(see Fig. 1). 
Hence there will be a net
spin current flowing without the accompanying charge current.
This is a {\it pure nonequilibrium spin current}.

We support our prediction by generalize
Landauer-B\"uttiker charge transport for multi-terminal spin transport.
We provide unambiguous definition of spin currents.
Using this theory we discusses the possibility of 
generating and detecting {\it nonequilibrium
pure spin currents} and point out the existence of {\it equilibrium
spin current}.

The possibility
of injecting pure spin currents were first discussed
in Ref. \cite{bauer1} for a three terminal device where two of the
terminals were ferromagnetic. 
Also direct optical injection of pure
spin currents in GaAs/AlGaAs quantum wells was demonstrated in
Ref. \cite{stev}. We would
stress that in our case spin current is not injected rather
generated intrinsically due to the spin-orbit interaction without
any magnetic
element in the system,
which is not the case in Ref.\cite{bauer1}.Thus we avoid the problem of spin injection 
altogether.  Further, since the effect
discussed relies on the general scattering properties due to the
spin-orbit interaction. Hence it will be observable with any kind of
spin-orbit
interaction, e.g.. Rashba spin-orbit interaction\cite{rashba}.

\begin{figure}
\begin{center}
\mbox{\epsfig{file=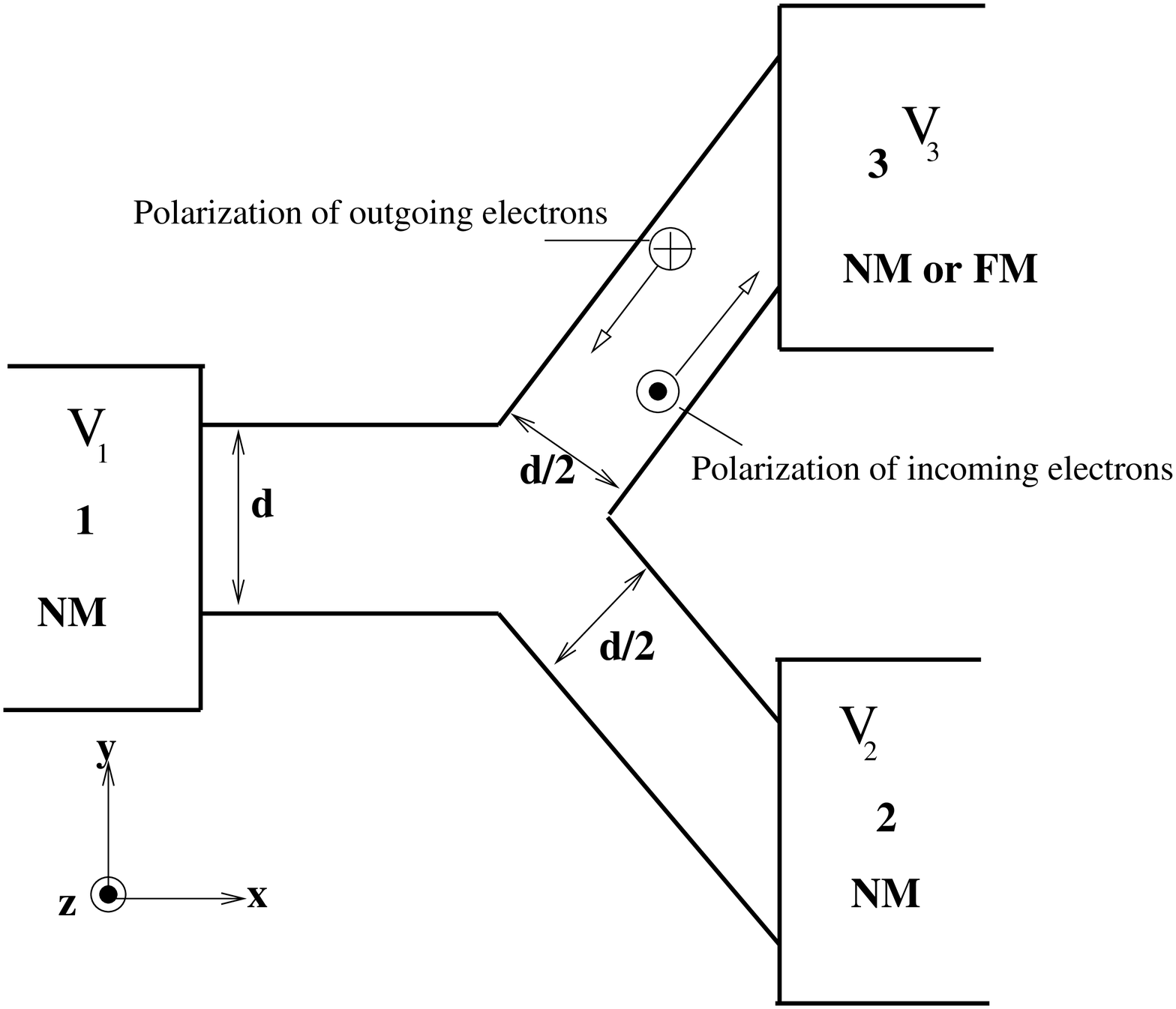,width=2.5in,height=2.5in,angle=0}}
\end{center}

\caption{Y-shaped three terminal junction with applied voltages
$V_{1}$, $V_{2}$ and $V_{3}$ as depicted. Terminal third (labeled 3)
is a voltage probe (non-magnetic or ferromagnetic) which 
draws no charge current. However the polarization incoming and
outgoing
electrons are opposite to each other, causing a {\it pure spin current}}
\label{Fig. 1}
\end{figure}

We first briefly outline the spin transport theory for multi-terminal
devices.
Let us consider the two dimensional Y-shaped
structure shown in Fig.1. The plane of structure is {\it xy} and a
perpendicular to it defines the coordinate system. Let us choose the
spin quantization axis to be along $\bf{\hat u}$, pointing along
($\theta$, $\phi$) where $\theta$ and
$\phi$ are usual spherical angles ( In other words we choose
the spin basis to be eigen states of operator  $\bbox{\sigma}$ $\cdot$
$\bf {\hat u}$ )
This is essentially since a charge current
flowing
along a spatially direction can be polarized along a direction 
which need not coincide with the direction of flow of charge current.
Also in presence of spin-orbit interaction the rotational invariance
in spin space is lost \cite{tribhu1} , hence any theory for spin
transport should take this fact into account.
With this definition we can generalize Landauer-B\"uttiker
theory for spin transport. Let { $V_{m}$} be potential at
terminal {\it m} measured from the minima of lowest band, where {\it m} can take values 1, 2 and 3
corresponding
to the three terminals of Fig. 1.
$T_{nm}^{\alpha, \sigma}$ is spin resolved
transmission probability of
electrons incident in lead {\it m} in spin channel
$\sigma$ to be transmitted
into lead {\it n} in spin channel $\alpha$.
We point out that $\sigma$ and $\alpha$
need not to be same in presence of spin-orbit interaction, since
SO interaction will make spin flip transmission probability non zero. The
spin current $\sigma$ flowing into terminal {\it m} is,
(here $\sigma$ can be either $\uparrow$ or $\downarrow$ )

\begin{equation}
I_{m}^{\sigma}=\frac{e^2}{h}\sum_{n \neq m, \alpha}(T_{n\, m}^{\alpha\,\sigma}V_{m}
-T_{m\, n}^{\sigma\, \alpha}V_{n})
\label{eq1}
\end{equation}

\noindent where $\alpha$ and $\sigma$ are indices labeling the
two spin eigenstates for a chosen quantization axis. 
In writing above equations
we have made an assumption that the spin
resolved transmission coefficient are energy independent. A
generalization of the above equation when the spin resolved
transmission
coefficient are energy dependent is straight forward. It amounts to
replacing $T_{n m}^{\alpha, \sigma}$ by $\int T_{n m}^{\alpha, \sigma}(E)$.

Since SO interaction preserves time reversal symmetry, which lead to
the following constrains on the spin-resolved transmission
coefficient,
\begin{equation}
T_{n\, m}^{\alpha\, \sigma} = T_{m\, n}^{-\sigma\, -\alpha}
\label{eq_sym}
\end{equation}

Using eq. 1 we can immediately write down 
the net charge and spin current flowing through terminal
{\it m},

\begin{equation}
I_{m}^{q}=I_{m}^{\sigma} + I_{m}^{-\sigma} \\
\equiv \frac{e^2}{h}\sum_{n \neq m, \sigma, \alpha}\{(T_{n\, m}^{\alpha\,\sigma}V_{m}
-T_{m\, n}^{\sigma\, \alpha}V_{n}\}
\label{eq_ch}
\end{equation}
  
\begin{eqnarray}
I_{m}^{s}&=&I_{m}^{\sigma} - I_{m}^{-\sigma}  \nonumber \\
&\equiv& \frac{e^2}{h}\sum_{n \neq m,  \alpha}\{(T_{n\, m}^{\alpha\,\sigma}- T_{n\, m}^{\alpha\,-\sigma})V_{m}
+(T_{m n}^{-\sigma\, \alpha}- T_{m n}^{\sigma\, \alpha}) V_{n}\}
\label{eq_sc}
\end{eqnarray}

\noindent where $I_{m}^{q}$ is charge current and $I_{m}^{s}$ is spin
current. We stress that eq. (\ref{eq_sc}) correctly determines spin
current generated by presence of spin orbit interaction. Since in the
absence of spin-orbit interaction and any magnetic element in the
device , spin resolved transmission
coefficient obey a further rotational symmetry in spin space {\it
  i.e.}
$T_{n\,m}^{\alpha, \sigma}=T_{n\,m}^{-\alpha,-\sigma}$, which implies
that spin currents are identically zero for all terminals, {\it i.e.},
$I_{m}^{s}=0$.

{\it Equilibrium spin current :} 
To discusses {\it equilibrium spin currents} let us consider
the case when all the potential are equal {\it i.e.}, $V_{m}=V_{0}$
$\forall$ m. In this situation charge current flowing in any terminal
should be zero ($I_{m}^{q}=0$)
which leads to the following sum rule (from eq. (\ref{eq_ch}))
\begin{equation}
\sum_{n}T_{n\, m}=\sum_{n}T_{m\, n}
\label{eq_sum}
\end{equation}
where $T_{n m}=\sum_{\alpha,\sigma}T_{n m}^{\alpha,\sigma}$ is total
transmission probability(summed over all spin channels) from terminal
{\it m} to {\it n}. This sum rule is robust and should be satisfied
irrespective of the detailed physics \cite{datta1}. This is a well
known
gauge invariance condition. Charge conservation  implies
$\sum_{m}I_{m}^{q}$=0 which follows from symmetry of spin resolved
transmission coefficient ,eq. (\ref{eq_sym}), and the gauge invariance
condition, eq.(\ref{eq_sum}). 
So in equilibrium there are
no charge currents flowing. However this is not the case for the spin
currents. This point can be appreciated if we look closely at the
equation (\ref{eq_sc}) for spin current.
Since in general the transmission coefficient,
 $ T_{n\, m}^{\alpha\,\sigma} \neq T_{n\,  m}^{\alpha\,-\sigma}$, which
 occurs in eq. (\ref{eq_sc}). Hence even when all the potential are
 equal , the spin current given by eq.(\ref{eq_sc})  is non zero. 
This is {\it equilibrium
spin current}.
Notice that this is consistent with  time reversal invariance
(eq. (\ref{eq_sym}))  
and the gauge invariance condition
given by eq. (\ref{eq_sum}). 
We would like to point out that this {\it equilibrium
spin current} would exist even in two terminal setup. 
The {\it equilibrium spin current} are carried by all the occupied
state
at a given temperature. This is a {\it non-linear
  response} and  is different from
linear response which give rise to {\it non-equilibrium spin currents}
and is a Fermi surface property. So strictly speaking for the
{\it equilibrium spin currents} one should take into account the
energy dependence of spin-resolved transmission coefficient.
A detailed study of the {\it equilibrium spin currents} would be
presented in a separate article \cite{tribhu2}. 
Here in this study we concentrate
more on the {\it non-equilibrium pure spin currents} and the related 
electrical effects.
Slonczewski has shown in Ref. \cite{sol} for magnetic multilayers
{\it equilibrium spin currents} causes non-local exchange coupling.
The important difference in our case is, we do not need Ferromagnetic contact to have
{\it equilibrium spin currents}, which was the case in Ref.\cite{sol}.
Rather in our case {\it equilibrium spin currents} are generated
intrinsically due to the  spin-orbit interaction without any magnetic
element in the system.

{\it Non-equilibrium spin currents:} To study {\it non-equilibrium
  spin currents}
,let us consider the case where the voltages at terminal 1
and 2 are respectively 
$V_1$=0 and $V_{2}$ and terminal third
is a voltage probe, i.e, $I_{3}^{q}=0$. With this condition one
can determine the voltage, $V_{3}$, at third terminal using the set of
equation (\ref{eq_ch}) and is given by \cite{tribhu3}, 
\begin{equation}
\frac{V_{3}}{V_{2}}=\frac{T_{32}}{T_{13}+T_{23}}
\label{volt}
\end{equation}
The spin current flowing through terminal 3 is
\begin{eqnarray}
\frac{I_{3}^{s}}={\frac{e^2}{h}}\sum_{\alpha}\left\{(T_{13}^{\alpha\,\sigma}-
T_{13}^{\alpha\,-\sigma}+T_{23}^{\alpha\,\sigma}-T_{23}^{\alpha\,-\sigma})V_{3}
\right. \nonumber \\ 
\left. +(T_{32}^{\alpha\,-\sigma}-T_{32}^{\alpha\,\sigma})V_{2} \right\} .
\label{I_3s}
\end{eqnarray}

From above equation (\ref{I_3s}) it is clear that $I_{3}^{s}$ is non-zero
while $I_{3}^{q}$ is zero by definition. Hence in 
terminal 3 there is a net spin currents flowing in the absence of any
net charge current. This is {\it pure spin current} and is
intrinsically generated by the spin-orbit interaction in the absence
of any magnetization as discussed in the
introduction.

To obtain quantitative results we perform numerical simulation on a
Y-shaped conductor shown in Fig. 1.
We model the conductor on a square tight binding lattice
with lattice spacing {\it a} and we use the corresponding tight
binding model including spin orbit interaction given by
eq.(\ref{eq_so}) \cite{tribhu}. For the calculation of spin resolved
transmission coefficient , we use the recursive green function
method. Details of this can be found in
Ref. \cite{tribhu},\cite{tribhu1}.
The numerical result presented are exact and takes the quantum effect
and multiple scattering into account.
For the model of disorder we take Anderson model,
where on-site energies are distributed randomly within [-U/2, U/2],
where U is the width of distribution. All the calculation were
performed on Y-shaped device of width {\it d=20a}, where {\it a} is
lattice spacing. Other parameters are given in figure captions.
\vspace{0.5cm}
\begin{figure}
\begin{center}
\mbox{\epsfig{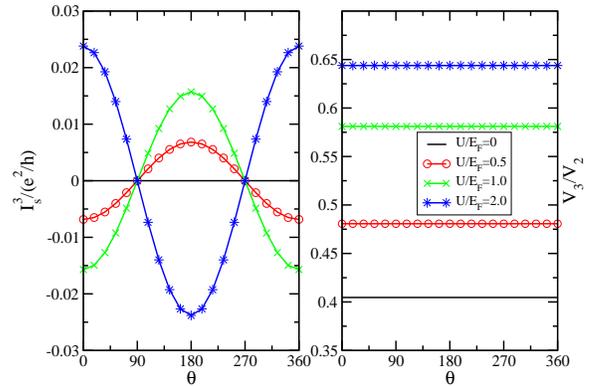}}
\end{center}
\caption{spin current given by (eq.(\ref{I_3s})) and potential given by (eq.(\ref{volt})) versus quantization axis for different values
  of disorder potential strength shown in inset.Calculations were
  performed on a device width $d=20a$ (see Fig.1) ,$k_{F}a=1$ ,
  dimensionless
spin-orbit
parameter $\lambda/a^{2}=0.05$}
\label{Fig.2}
\end{figure}

In Fig.\ref{Fig.2} we show the spin currents $I_{s}^{3}$ flowing through the
terminal 3 (Right panel) and the corresponding voltage $V_{3}$ when
all the three terminal are non-magnetic.
In Fig.\ref{Fig.2} , $\theta$=0
corresponds
to {\it z axis} and $\theta$=90 corresponds to {\it y axis}, we have
kept fixed $\phi=90$.
We see
that the maximum amount of spin currents flow along {\it z axis}. This
is understandable since for strictly two dimensional case , the
spin-orbit
coupling given by eq. (\ref{eq_so}) conserves {\it z} component of
spin.
Hence the asymmetric scattering produced by spin orbit interaction
causes a pure spin currents along {\it z} axis, as discussed in
introduction. For the ballistic case (curve for $U/E_{F}=0$), spin currents is
zero since there is no spin-orbit interaction in this case, as can be
seen from eq. (\ref{eq_so}) by putting the potential $U(r)=0$.
Also for strong disorder spin current changes sign 
(curve for $U/E_{F}=2$) due to multiple
scattering.
We would like to stress that by definition charge current flowing
in terminal 3 is zero, {\it i.e.} $I_{3}^{q}=0$.  Now from the right
panel in Fig.\ref{Fig.2}  we see that the voltage $V_3$ measured is different
although
there is no charge current flowing and the magnitude of $V_3$ is
directly proportional to the {\it z} component of spin current. 
As is seen , with the increase of disorder strength the magnitude of
spin currents increases and accordingly the potential $V_3$ also
increases. However potential $V_3$ is independent of quantization axis
since the voltage probe is non-magnetic. {\it Hence with a non-magnetic
voltage probe one can detect the spin current , but can not measure
it}.  To measure the spin currents one would need a ferromagnetic
voltage probe. An intuitive understanding of this can be gained
as follows. From Fig.\ref{Fig.2} (right panel) we notice that the spin currents
depends on the quantization axis. Thus if the probe is a
ferromagnetic, electrons which are polarized
parallel to the ferromagnet would be transmitted easily than the
electrons polarized anti-parallel to the ferromagnet.Since the
voltage at the probe is determined by the ration of transmission
coefficient
(eq.(\ref{volt})), hence the probe voltage 
should show variation in
phase with the spin currents.

\begin{figure}
\begin{center}
\mbox{\epsfig{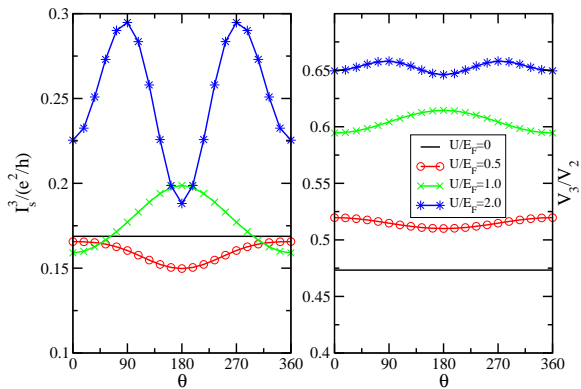}}
\end{center}
\caption{spin current flowing through the Ferromagnetic terminal 3
(voltage probe)
versus quantization axis (left panel) and right panel shows the
corresponding
voltage at the terminal third.Different curves corresponds to
different strength of disorder. Inset shows the strength of Anderson
disorder. Ferromagnet is
modeled as exchange split with exchange splitting ($\Delta$) given as
$\Delta/E_{F}=0.5$. Other parameters are same as for Fig. 2. }
\label{Fig.3}
\end{figure}

This is confirmed in Fig.\ref{Fig.3}. Where we
have plotted spin current (left panel) and voltage (right panel) for
the case when the third terminal is a ferromagnetic. 
Left panel shows the spin currents and the corresponding voltage is
shown in right panel. The quantization axis is given by the
direction of magnetization. We see that as the spin current changes the
corresponding voltage measured also changes in phase. Hence by having
a ferromagnetic voltage probe one can measure the {\it pure spin
  current}. We would like to mention that in our numerical simulation
voltage probe is an invasive one, {\it i.e.} it is strongly coupled to
the system, hence one sees the quantitative difference between the
results of Fig.\ref{Fig.2}  and Fig.\ref{Fig.3}. In Fig.\ref{Fig.3} , we see that spin currents are
non
zero for the ballistic case ($(U/E_{F}=0)$) while for non-magnetic
case shown in Fig.\ref{Fig.2}  spin current for ballistic case is zero.
This is so because
the Ferromagnetic probe is strongly coupled(invasive probe), it essentially injects
a polarized current.  
However this is not a hindrance for measuring
spin currents generated by spin-orbit interaction. Since as is seen
from Fig.\ref{Fig.3} , it only gives rise to a constant shift
compared to the non-magnetic case (Fig\ref{Fig.2}).
Recently it was pointed out in ref. \cite{bauer} that in magnetic bilayer systems
dynamic exchange coupling arises due to the injection of polarized
charge current.
In the said work effect of spin-orbit interaction were not taken into
account. Since here we point out the existence of spin
currents (equilibrium and nonequilibrium) due
to the spin-orbit interaction, hence such currents in principle would modify 
quantatively proposed dynamic exchange coupling.
Since charge transport for the Y-shaped mesoscopic junction have been
studied in past experimentally as well theoretically. In view of this
we hope the study presented here for the spin transport would open up
new
opportunity in the field of spintronics.

Author acknowledges fruitful discussion with P. Bruno, G. Bouzerar, Y.
Utsumi and A. M. Jayannavar. Author also
acknowledge discussion with 
Gerrit. E. W. Bauer and for pointing out Ref.\cite{sol}.

This work was financially supported by the German Federal Ministry of
Research (BMBF)

\vspace{0.5cm}

\begin{thebibliography}{}
\bibitem{wolf}S. A. Wolf,D. D. Awschalom, R. A. Buhram,
  J. M. Daughton,S. Von. et. al., Science {\bf 294}, 1488 (2001) and
  references therein.
\bibitem{molenkamp} L. W. Molenkamp {\it et al.}, Phys. Rev. B. {\bf
    64}, R121202 (2001).
\bibitem{datta} S. Datta and B. Das, Appl. Phys. Lett.{\bf 56},
  665(1990).
\bibitem{hirsch} J. E. Hirsch, Phys. Rev. Lett. {\bf 83}, 1834 (1999),
Shufeng Zhang, Phys. Rev. Lett. {\bf 85}, 393 (2000).
\bibitem{bergman} G. Bergman, Phys. Rev. B. {\bf 63}, 193101-1 (2001).
\bibitem{tribhu}  T. P. Pareek and P. Bruno, Phys. Rev. B.{\bf 63}, 165424-1
(2001). P. Bruno Phys. Rev. Lett. {\bf 79}, 4593 (1997). 
\bibitem{Landau} L.D. Landau and E. M. Lifshitz {\it Quantum Mechanics
Vol. 3}, pp. 583-588.

\bibitem{kiseleve} A. A. Kiseleve  and
K. W. Kim cond-mat 0203261.

\bibitem{but1} M. B\"uttiker, IBM J. Res. Develop. {\bf 32} 63 (1988).
\bibitem{tribhu3} T. P. Pareek, Sandeep K. Joshi and A. M. Jayannavar
{\bf 57}, 8809 (1998).
\bibitem{rashba} Yu. A. Bychkov and E. I. Rashba, Sov. Phys. JETP Lett.
{\bf 39}, 78 (1984).
\bibitem{bauer1} A. Brataas, Yu. V. Nazarov and G.E.W. Bauer,
  Phys. Rev. Lett {\bf 84}, 2481 (2000).
\bibitem{stev} M. J. Stevens, A. L. Smirl, R. D. R. Bhat, A. Najmaie,
  J. E. Sipe,and H. M. van. Driel, Phys. Rev. Lett. {\bf 90}, 136603-1
  (2003).
\bibitem{tribhu1} T. P. Pareek, Phys. Rev. B. {\bf 66}, 193301 (2002).
T. P. Pareek and P. Bruno, Phys. Rev. B. {\bf 65}, 241305 (2002).
\bibitem{datta1} S. Datta,{\it Electronic transport in mesoscopic
systems}, (Cambridge University press, Cambridge, 1997). 
\bibitem{tribhu2} T. P. Pareek, manuscript under preparation.
\bibitem{sol} J.~C.~Slonczewski, Journal of Magnetism and Magnetic
  Materials.
{\bf 126}, 374 (1993).
\bibitem{bauer} B. Heinrich, Y. Tserkovnyak, G. Woltersdorf,
  A. Brataas, R. Urban and Gerrit E. W. Bauer, Phys. Rev. Lett. {\bf
    90},
187601-1 (2003)

\end{thebibliography}
\end{document}